\documentclass[12pt]{article}
\usepackage{amssymb,amsmath,cite}
\usepackage{epsfig}
\long\def\@makefntext#1{\parindent 0cm\noindent \hbox to
1em{\hss$^{\@thefnmark}$}#1}

\begin{document}
\begin{titlepage}
\vspace{.5in}
\begin{flushright}
\end{flushright}
\vspace{.5in}
\begin{center}
{\Large\bf  Possible dark energy imprints in gravitational wave spectrum  of mixed neutron-dark-energy stars}\\
\vspace{.4in} {Stoytcho~ S.~Yazadjiev$^{1,2}$\footnote{\it email: yazad@phys.uni-sofia.bg}, Daniela D.
Doneva$^{2}$\footnote{\it
email: doneva@tat.physik.uni-tuebingen.de}\\
       {\footnotesize\it ${}^{1}$ \it Department of Theoretical Physics, Faculty of Physics,}
       {\footnotesize \it Sofia University, Sofia, 1164, Bulgaria }\\
            { \footnotesize  ${}^{2}$ \it Theoretical Astrophysics, Eberhard-Karls University of T\"ubingen, T\"ubingen 72076, Germany }}
\end{center}

\vspace{.5in}
\begin{center}
{\large\bf Abstract}
\end{center}
In the present paper we study the oscillation spectrum of neutron
stars containing both ordinary matter  and dark energy in different
proportions. Within the model we consider, the equilibrium
configurations  are numerically constructed and the results show
that the properties of the mixed neuron-dark-energy star can differ
significantly when the amount of dark energy in the stars is varied.
The oscillations of the mixed neuron-dark-energy stars are studied
in the  Cowling approximation. As a result we find that the
frequencies of the fundamental mode and the higher overtones are
strongly affected by the  dark energy content. This can be used in
the future to detect the presence of dark energy in the neutron
stars and to constrain the dark-energy models.
\end{titlepage}
\addtocounter{footnote}{-1}

\section{Introduction}
The cosmological  observations of the present Universe provide
evidences for the existence of a mysterious kind of matter, called
dark energy, which governs the accelerated expansion of the Universe
\cite{R1},\cite{ASSS}. The dark energy constitutes $73\%$ of total
energy content of the Universe and it exhibits some unusual
properties such as negative pressure to density ratio $w$ (in
hydrodynamical language).

The fundamental role that the dark energy plays in  cosmology
naturally makes us  search for local astrophysical manifestation of
it. If dark energy was discovered at astrophysical scales it would
give us new tools for more profound investigation of it and its
properties. For example, the existence of dark energy makes us
expect that some of the present-day stars are a mixture of both
ordinary matter and dark energy in different proportions. The study
of such mixed objects is a new interesting problem in the current
investigations, see for example \cite{MM}--\cite{Y} and references
therein.

The natural question that arises is how we  can detect the presence
of dark energy in the stars. Are there any imprints of dark energy
that can be observed and how? In present paper we consider one such
possibility based on the gravitational waves radiation by the
neutron stars. It is well known that the characteristics,  and more
precisely the spectrum of the gravitational waves emitted by the
compact objects, depend on the interior  structure of the compact
objects. This is the base of the asteroseismology -- the interior
structure of the compact objects can be revealed by investigating
the spectrum of the gravitational waves they emit
\cite{Kokkotas01,Andersson98}. Currently a lot of efforts are
devoted to the detection of the gravitational waves. Several
ground-based detectors as LIGO, VIGRO, TAMA300 and GEO600 are
operating and the launching of space-based detectors, for example
LISA, are expected in near future. The detection of the
gravitational waves will open a new  window to the Universe and will
provide us with a unprecedented tool for studying the compact
astrophysical objects.

In the present paper we consider a certain model of mixed
neutron-dark-energy stars (MNDES) and study their oscillation
spectrum in the Cowling approximation. We show that the spectrum of
MNDES contains imprints  of dark energy and could be used to detect
the presence of dark energy in the neutron stars.

\section{Equilibrium model for mixed neutron-dark-energy stars}

The nature of the dark energy is a mystery at present and it is a
great challenge to current physics to solve this problem.
Nevertheless, in their attempts to understand and model the dark
energy, theoretical physicists have adopted some effective
descriptions of it. The most simple and direct description is by a
perfect fluid with a negative pressure. The other effective
descriptions of the dark energy depend on the value of $w$. If
$w>-1$ a possible description is provided by a scalar field with an
appropriately chosen potential. In the case when $w=-1$ the dark
energy can be explained by the cosmological constant. If $w<-1$ then
a possible description of dark energy  is provided by scalar fields
with negative kinetic energy, the so-called phantom (ghost) scalars
\cite{Caldwell},\cite{CSSX}. The negative kinetic energy, however,
leads to severe quantum instabilities\footnote{The perfect fluid
description of the dark energy also suffers from instabilities due
to the imaginary velocity of the sound. From a classical point of
view the massless phantom field is even more stable than  its usual
counterpart \cite{BCCF},\cite{APicon}.  } and this is a formidable
challenge to the theory. However, there are claims that these
instabilities can be avoided \cite{PT}.  In general, the problem
could be avoided if we consider the phantom scalars as an effective
field theory resulting from some kind of fundamental theory with a
positive energy \cite{NO},\cite{CHT}. In this case the phantom
scalar description of the dark energy is physically acceptable. In
this context it is worth noting that the phantom-type fields arise
in string theories and  supergravity \cite{Sen1}--\cite{Nilles}.

At present  it is not clear whether $w<-1$, $w=-1$ or $w>-1$. It seems, however, that the current observational
constraints favor $w<-1$ \cite{Percival}--\cite{Li}. That is why in the present work we adopt an effective description
of the dark energy by a phantom scalar. Since a very little is known about the interaction of dark energy with the
normal matter we consider  a simple model with minimal interactions  for the phantom field -- we do not  include terms
describing non-minimal interaction between the phantom field and the normal matter in the field equations \cite{Y}.
Under these conditions the Einstein equations in the presence of dark energy read \cite{Y}
\begin{eqnarray}\label{FE}
&&R_{\mu\nu}= 8\pi (T_{\mu\nu} - {1\over 2}Tg_{\mu\nu}) - 2\partial_{\mu}\varphi\partial_{\nu}\varphi,\\
&&\nabla_{\mu}\nabla^{\mu}\varphi= 4\pi \rho_{D}. \nonumber
\end{eqnarray}
Here $T_{\mu\nu}$ is the energy-momentum tensor of the ordinary
matter in the perfect fluid description with energy density $\rho$
and pressure $p$:
\begin{eqnarray}
T_{\mu\nu}=(\rho + p)u_{\mu}u_{\nu} + pg_{\mu\nu}.
\end{eqnarray}
For the ordinary matter we impose the  natural conditions $\rho\ge
p\ge 0$.   The density of the dark energy sources is denoted by
$\rho_{D}$. The dark energy sources will be called {\it dark
charges}. When we consider the local manifestation of  dark energy
on astrophysical scales, i.e. scales much smaller than the
cosmological scales, the phantom potential ${\cal U}(\varphi)$ can
be neglected and that is why we have set ${\cal U}(\varphi)=0$.

The contracted Bianchi identity applied to the field equations gives

\begin{eqnarray}\label{CEMT}
\nabla_{\nu}T^{\nu}_{\mu}= \rho_D \nabla_{\mu}\varphi.
\end{eqnarray}
This equation projected orthogonally to the 4-velocity $u^{\mu}$
gives the equation of motion of the fluid. In the case of static
configurations  (not necessary spherically symmetric) it reduces to
the equation describing the hydrostatic equilibrium:

\begin{eqnarray}\label{SHEE}
(\rho+p)\partial_{i}\ln\sqrt{|g_{tt}|} + \partial_i p = \rho_{D}\partial_{i}\varphi,
\end{eqnarray}
where the subscript $i$ is for the spatial coordinates. For barotropic equation of state (EOS) of the ordinary  matter
$p=p(\rho)$, eq. (\ref{CEMT}) can be rewritten in the form

\begin{eqnarray}
d\ln\sqrt{|g_{tt}|} + d\int \frac{dp}{\rho + p} = \frac{\rho_D}{\rho
+ p}d\varphi.
\end{eqnarray}
The integrability condition for this equation gives
\begin{eqnarray}\label{EOSDE}
\rho_{D}= f(\varphi)(\rho + p),
\end{eqnarray}
where $f(\varphi)$ is  arbitrary function of $\varphi$. In the
present paper we consider the simplest choice for $f(\varphi)$,
namely $f(\varphi)=k$ where $k\ge 0$  is a constant, i. e.

\begin{eqnarray}\label{RDES}
\rho_{D}= k (\rho+ p).
\end{eqnarray}

In fact the constant $k$ is a measure for the content of dark energy
in the star, while $(\rho+p)$ determines the dark energy profile in
the star.

Now let us consider the projection of (\ref{CEMT}) along the
4-velocity $u^\mu$. We find

\begin{eqnarray}
-\frac{d\rho}{d\tau} - (\rho +p)\nabla_{\nu}u^{\nu}= \rho_D
\frac{d\varphi}{d\tau},
\end{eqnarray}
where $\frac{d}{d\tau}$ is the derivative along the fluid
4-velocity. Combining this equation with the baryon conservation law
$\nabla_{\nu}\left(n u^{\nu}\right)=0$ we find

\begin{eqnarray}
-\frac{d\rho}{d\tau} + \frac{(\rho +p)}{n} \frac{dn}{d\tau}=\rho_D
\frac{d\varphi}{d\tau},
\end{eqnarray}
or equivalently

\begin{eqnarray}\label{FTL}
d\rho=  \frac{(\rho +p)}{n} dn - \rho_D d\varphi,
\end{eqnarray}
which is the  thermodynamics first  law for our system of ordinary
matter and dark energy. Hence we conclude that

\begin{eqnarray}
\rho=\rho(n,\varphi),
\end{eqnarray}
i.e. the energy density is a function not only of the particle
number density but depends also on the scalar field. In order to
find the function $\rho(n,\varphi)$ we use the barotropic equation
$p=p(\rho)$ and eq. (\ref{RDES}) which substituted in (\ref{FTL})
and some algebra gives

\begin{eqnarray}\label{EFRHO}
\int_{0}^{\rho} \frac{d\rho}{\rho + p}=\int^{n_{*}}_{0}
\frac{dn_{*}}{n_*},
\end{eqnarray}
where

\begin{eqnarray} n_{*}= e^{-k\varphi} n.
\end{eqnarray}
Eq. (\ref{EFRHO}) determines the function $\rho(n,\varphi)$ and it
is easy to see that if in the absence of dark energy $\rho(n)=F(n)$,
then $\rho(n,\varphi)= F(n_*)=F(e^{-k\varphi} n)$ in the presence of
dark energy.

From now on we will focus on spherically symmetric equilibrium
solutions. In this case the spacetime metric can be written in the
well-known form

\begin{eqnarray}
ds^2= - e^{2\Phi(r)}dt^2 + e^{2\Lambda(r)}dr^2 + r^2(d\theta^2 +
\sin^2\theta d\phi^2)
\end{eqnarray}
and the reduced field equations are the following
\begin{eqnarray}\label{RFE}
&&\frac{d}{dr}\left[r\left(1- e^{-2\Lambda}\right)\right]=8\pi \rho
r^2 - r^2 \left(\frac{d\varphi}{dr}\right)^2 e^{-2\Lambda}, \nonumber \\
&&\frac{d\Phi}{dr}= \frac{1}{2r} \left[e^{2\Lambda} -1 + 8\pi pr^2
e^{2\Lambda} - r^2 \left(\frac{d\varphi}{dr}\right)^2 \right], \\
&&\frac{d}{dr}\left[e^{\Phi-\Lambda} r^2 \frac{d\varphi}{dr}
\right]= 4\pi \rho_{D}r^2 e^{\Phi + \Lambda}, \nonumber
\end{eqnarray}
together with the equation describing hydrostatic equilibrium
(\ref{SHEE}) in the spherically symmetric case

\begin{eqnarray}\label{SSSE}
(\rho+p)\frac{d\Phi}{dr} + \frac{dp}{dr}=
\rho_{D}\frac{d\varphi}{dr}.
\end{eqnarray}

Here it is worth noting that the system of equations
(\ref{RFE})--(\ref{SSSE}),  can be viewed as describing an
equilibrium spherically symmetric configuration with an effective
anisotropic pressure\footnote{ For a discussion of locally
anisotropic selfgravitating systems we refer the reader to
\cite{Herrera}.}. As a hole we can say that the presence of dark
energy in the star yields a pressure anisotropy of certain kind and
(\ref{RDES}) can be considered as an effective quasi-local equation
of state for the anisotropic part of the pressure. The local
anisotropy also affects the thermodynamics first law as we have
already seen in our case.

We numerically solve the system (\ref{RFE})--(\ref{SSSE}) with the following boundary conditions

\begin{eqnarray}
\Lambda(0)=0,\; \rho(0)=\rho_0, \; \frac{d\varphi}{dr}(0)=0, \;
\varphi(\infty)=0, \Phi(\infty)=0.
\end{eqnarray}

The radius $R$ of the star is determined form the condition $p(R)=0$. We also use a polytropic equations of state for
the ordinary matter

\begin{eqnarray}
&&p=p(n_*)= K {n}_0 m_b \left(\frac{n_*}{{n}_0}\right)^\Gamma= K {n}_0 m_b \left(e^{-k\varphi}\frac{n}{{n}_0}\right)^\Gamma, \\
&&\rho=\rho(n_*)=n_{*} m_b + \frac{p}{\Gamma-1}= e^{-k\varphi} n m_b
+ \frac{p}{\Gamma-1},
\end{eqnarray}
where $n$ is the particle number density, $m_b=1.66 \times 10^{-24} g$, $n_0=0.1 fm^{-3}$. We have chosen $\Gamma=2.46$
and $K=0.00936$ obtained when fitting the tabulated data for  EOS II \cite{Alonso85}, but the results are qualitatively
the same for other values $K$ and $\Gamma$.

The mass, the number of baryons  and the dark charge of the star are
given by \cite{Y}

\begin{eqnarray}
M&=&\! -{1\over 4\pi}\int_{Star} R_{t}^{t}\sqrt{-g}d^3x =\int_{Star}
(\rho + 3p)\sqrt{-g}d^3x = 4 \pi \int_0^R {(\rho + 3p)\, e^{\Phi + \Lambda}r^2 dr} , \\\nonumber \\
N&=&4\pi  \int_0^R n e^{\Lambda} r^2 dr,
\\
 D&=&\int_{Star} \rho_{D}\sqrt{-g}d^3x =4\pi  \int_0^R \rho_D
\,e^{\Phi + \Lambda}r^2 dr = {1\over 4\pi} \oint_{S_{\infty}^2}
\nabla_{\mu}\varphi d\Sigma^{\mu}.
\end{eqnarray}
Other equivalent formulas for calculating $M$ and $D$ can be
obtained  by using eqs. (\ref{RFE}).

\begin{figure}
\includegraphics[width=0.45\textwidth]{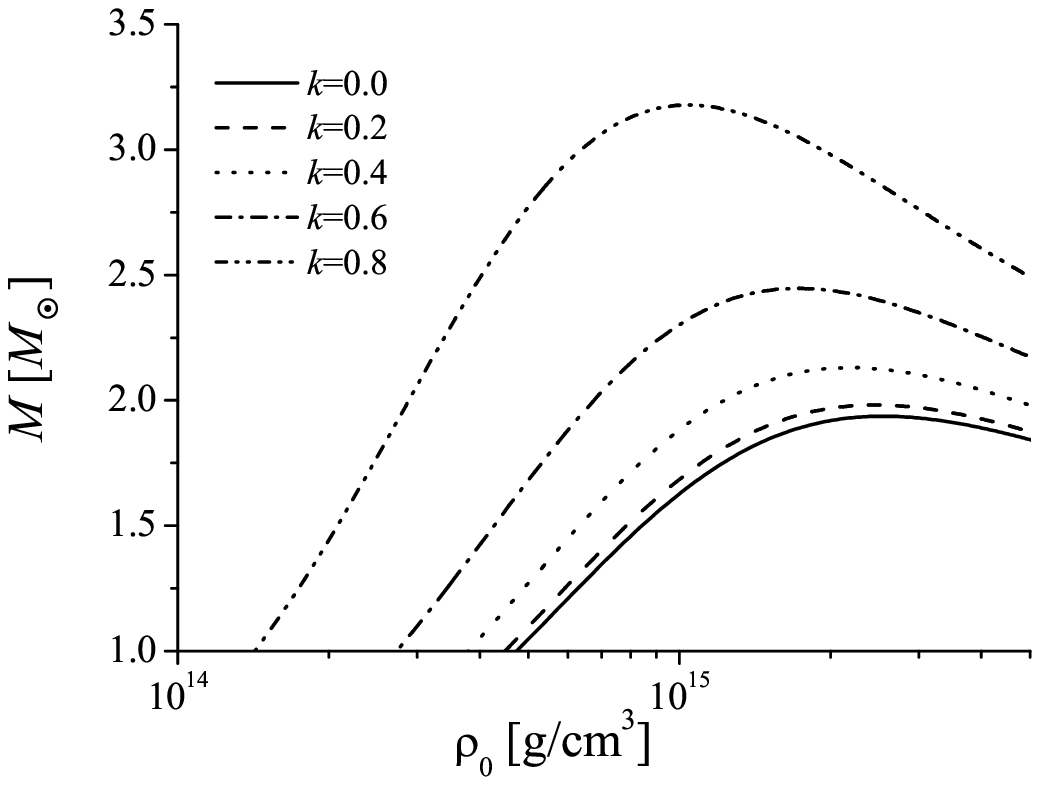}
\includegraphics[width=0.45\textwidth]{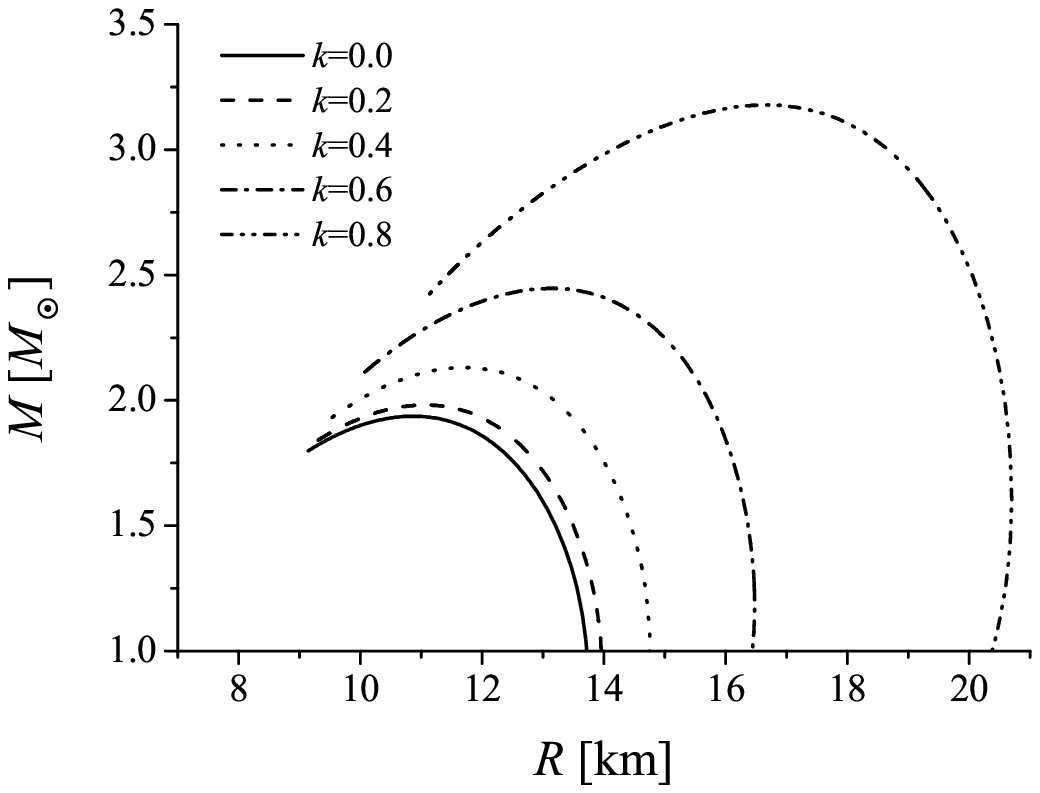}
\caption{The mass of the MNDES as a function of the central density (left panel) and of radius of the ordinary matter
component (right panel) for several values of the dark energy parameter $k$ ($k=0$ corresponds to a pure neutron star).} \label{fig:MRrho_Background}%
\end{figure}

\begin{figure}
\includegraphics[width=0.48\textwidth]{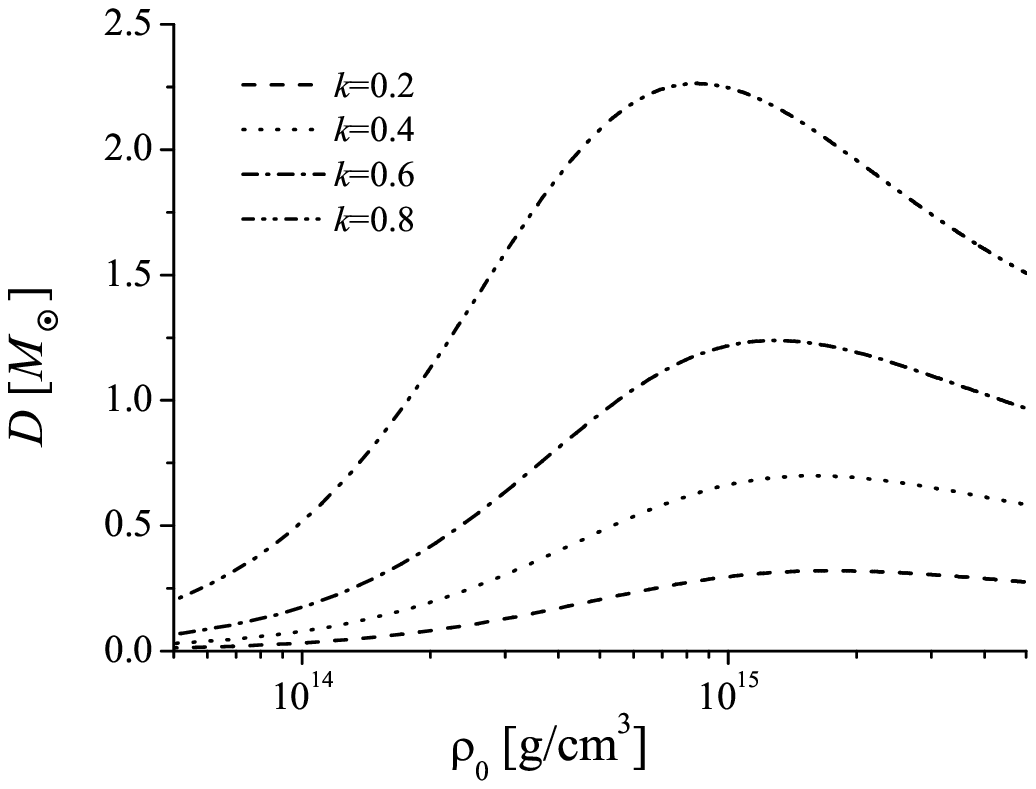}
\includegraphics[width=0.48\textwidth]{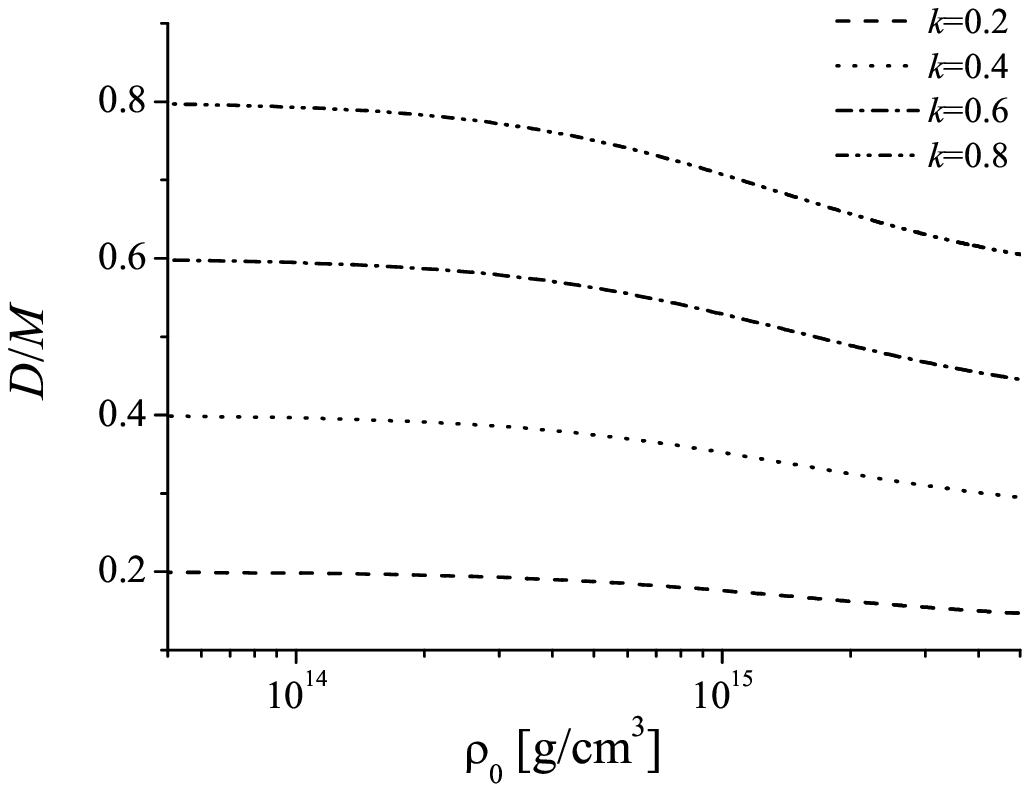}
\caption{The dark charge in solar mass units (left panel) and the ratio of the dark charge to the total mass of the
MNDES (right panel) as functions of the central density. The results are for the same solutions as those shown on
Fig. \ref{fig:MRrho_Background}.} \label{fig:DM(rho)_Background}%
\end{figure}

The numerical results for the background solutions are shown on Fig.
\ref{fig:MRrho_Background} where the mass $M$ as a function of the
central density $\rho_0$ and of the radius $R$ are presented. On
Fig. \ref{fig:DM(rho)_Background} the dark charge $D$ and the ratio
of the dark charge to the total mass $D/M$ of the MNDES as functions
of the central density are shown. The sequences of solutions are
obtained when varying the value of the cental density of the
ordinary matter $\rho_0$ while keeping the dark energy parameter $k$
fixed. As we can see the properties of the star can vary
significantly when the portion of the dark energy inside the star
increases, i.e. when we increase $k$. For large values of $k$ (i.e.
large portions of dark energy) the total mass of the star can be
much larger than the pure neutron star mass for the same equation of
state and the radius of the star also increases. These effects are
due to the phantom field which yields repulsive rather than
attractive force. In fact the additional force on the right hand
side of eq.(\ref{SSSE}) is repulsive, $\rho_D\frac{d\varphi}{dr}>0$
and this can be easily seen form the field equation for $\varphi$.
In hydrodynamical language, the additional force due to the
anisotropic nature of the fluid is pointing outward. We have found
numerically that there are no equilibrium configurations for
$k\gtrsim 1$ since the repulsive force dominates on the attractive
gravity in that case.

\begin{figure}
\begin{center}
\includegraphics[width=0.6\textwidth]{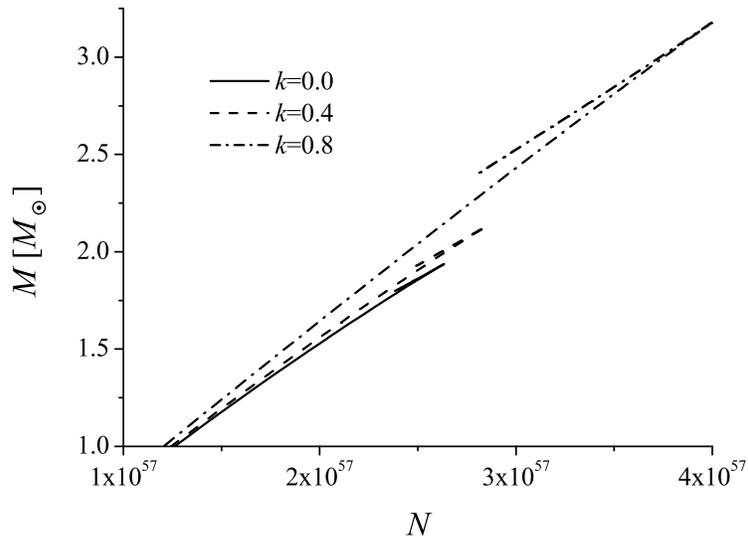}
\end{center}
\caption{The dependence of the star mass on the ordinary matter
particle number for different values of $k$.} \label{fig:M(N)_Background}%
\end{figure}

Now let us turn to  the question about the stability of the
equilibrium configurations. The dynamical stability is usually
studied by considering their radial perturbations and reducing the
problem to an eigenvalue problem. Since the equilibrium
configurations in our case are parameterized by one parameter (for
fixed $k$), say the central density of the ordinary matter
$\rho_{0}$, the eigenvalues will be continuous functions of them.
What is important in the present context is the point between the
stable and instable configurations. On that point we must have a
zero eigenvalue or in other words a static perturbation transforming
the equilibrium state  with $\rho_{0}$ to the equilibrium state with
$\rho_{0}+ \delta \rho_{0}$. The static perturbation preserves the
number of particles $N$ of the ordinary matter.  Therefore the
perturbed equilibrium state have the same $N$ as the initial
equilibrium state which means that

\begin{eqnarray}
\delta N= \frac{\partial N}{\partial \rho_{0}}\delta\rho_{0}=0.
\end{eqnarray}
Therefore the change of stability is where $\frac{\partial
N}{\partial \rho_0}=0$. Moreover, for our model and for fixed $k$,
one can show that\footnote{More precisely we have $\delta M= {\tilde
\mu} \delta N  + D\delta\varphi_{\infty}$, however we consider fixed
cosmological value  $\varphi_{\infty}$.}
\begin{eqnarray}
\delta M= {\tilde \mu} \delta N,
\end{eqnarray}
where ${\tilde \mu}$ is the red-shifted chemical potential, ${\tilde
\mu}= \frac{\rho+ p}{n}e^{\Phi}$. It can be shown from (\ref{SSSE})
that ${\tilde \mu}$ is constant throughout the star. Consequently
the mass and the particle number peak at the same location, as one
can see form Fig. \ref{fig:M(N)_Background}. Therefore the stability
change occurs at the critical energy density $\rho^c_0$ for which
$\frac{\partial M}{\partial \rho_{0}}=0$. The configurations with
$\rho_0<\rho^c_0$ are stable while those with $\rho_0>\rho^c_0$ are
instable because the configurations with $\rho_0<\rho^c_0$ have
lower energy than those with $\rho_0>\rho^c_0$ as one sees on Fig.
\ref{fig:M(N)_Background}.

\section{Perturbation equations in Cowling approximation}

In this section we derive the equations describing the non-radial
perturbations of the MNDES in the so-called Cowling approximation
\cite{McDermott}--\cite{Lindblom}. In Cowling approximation the
non-fluid degrees of freedom are kept fixed -- for example, within
the framework of general relativity, the spacetime metric is kept
fixed. Despite of this simplification the Cowling formalism turns
out to be accurate enough and reproduces the oscillation spectrum
with good accuracy. In fact the comparison of the oscillation
frequencies obtained by a fully general relativistic numerical
approach and by the Cowling approximation shows that the discrepancy
is less than $20 \%$ for the typical stellar models \cite{Yoshida}.
The Cowling approximation is trustable even in the case of presence
of scalar fields \cite{Sotani04}. We apply the Cowling approximation
to our model by keeping the metric and the scalar phantom field
frozen.

The equations describing the perturbations in Cowling formalism are
obtained by varying the equation for the conservation of the
energy-momentum tensor (\ref{CEMT}). Taking into account that the
metric and scalar field are frozen, we find $\nabla_{\nu}\delta
T^{\nu}_{\mu}=0$ where

\begin{eqnarray}
\delta T^\nu_{\mu}= \left(\delta \rho + \delta p\right) u_{\mu}
u^{\nu} + \left(\rho + p\right)\left(u_{\mu}\delta u^{\nu} + \delta
u_{\mu} u^{\nu} \right).
\end{eqnarray}

Projecting the equation $\nabla_{\nu}\delta T^{\nu}_{\mu}=0$  along
the background 4-velocity $u^{\mu}$  we have

\begin{eqnarray}
u^{\nu}\nabla_{\nu}\delta\rho + \nabla_{\nu}\left[(\rho + p)\delta
u^{\nu}\right] + \left(\rho + p\right)a_{\nu}\delta u^{\nu}=0.
\end{eqnarray}

Projecting orthogonally to the background 4-velocity by using the
operator $P_{\mu}^{\nu}=\delta^{\nu}_{\mu} + u^{\nu}u_{\mu}$, we
obtain

\begin{eqnarray}\label{DPE}
\left(\delta\rho + \delta p\right)a_{\mu} + \left(\rho + p\right)u^{\nu} \left(\nabla_{\nu}\delta u_{\mu} -
\nabla_{\mu}\delta u_{\nu} \right) + \nabla_{\mu}\delta p + u_{\mu} u^{\nu}\nabla_{\nu}\delta p=0,
\end{eqnarray}
where $a_{\mu}=u^{\nu}\nabla_{\nu}u_{\mu}$ is the background
4-acceleration. Taking into account that $u=(u^t, 0,0,0)$ the above
equation can be rewritten in the form
\begin{eqnarray}\label{POE}
\left(\delta\rho + \delta p\right)a_{i} + \left(\rho +
p\right)u^{t}\partial_{t}\delta u_{i} + \partial_{i}\delta p=0
\end{eqnarray}
with $i=1,2,3=r,\theta,\phi$

At this stage we can express the perturbations of the 4-velocity via
the Lagrangian displacement vector $\xi^{i}$, namely

\begin{eqnarray}
\frac{\partial \xi^{i}}{\partial t}= \frac{\delta u^{i}}{u^t}.
\end{eqnarray}

Now let us consider  eq.(\ref{POE}) for $i=1$ and $i=2$. Since
$a_{\theta}=a_{\phi}=0$ we find

\begin{eqnarray}
(\rho + p)(u^{t})^2\partial^2_{t}\xi_{\theta}+
\partial_{\theta}\delta p=0, \\ \nonumber \\
(\rho + p)(u^{t})^2\partial^2_{t}\xi_{\phi}+
\partial_{\phi}\delta p=0.
\end{eqnarray}
Taking into account that $\rho, p$ and $u^{t}$ depend on $r$ only,
the integrability condition for the above equations gives

\begin{eqnarray}
\partial_{\theta}\xi_{\phi}= \partial_{\phi}\xi_{\theta}.
\end{eqnarray}
From this condition and the fact that the background is spherically
symmetric we find that $\xi_{\theta}$ and $\xi_{\phi}$ are of the
form

\begin{eqnarray}
\xi_{\theta}= -\sum_{lm}V_{lm}(r,t)\partial_{\theta}Y_{lm}(\theta,\phi),\\ \nonumber\\
\xi_{\phi}= -\sum_{lm}V_{lm}(r,t)\partial_{\phi}Y_{lm}(\theta,\phi),
\end{eqnarray}
where $Y_{lm}(\theta,\phi)$ are the spherical harmonics. From now on, in order to simplify the notations, we will just
write $\xi=-V Y_{lm}$ when we have expansion in spherical harmonics.

We proceed further with finding the expressions for the density and
pressure  perturbations. From eq.(\ref{DPE}) we have

\begin{eqnarray}
u^{t}\delta \rho = -
\frac{1}{\sqrt{-g}}\partial_{i}\left[\sqrt{-g}(\rho
+p)u^{t}\delta\xi^{i}\right] - (\rho + p)a_{i}\xi^{i}.
\end{eqnarray}
It is convenient to express $\xi^{r}$ in the form
\begin{eqnarray}
\xi^{r}= e^{-\Lambda} \frac{W}{r^2} Y_{lm}
\end{eqnarray}
 and substituting in the above equations, after some algebra
we find\footnote{The derivative with respect to the radial
coordinate $r$ will be denoted by prime or by the standard symbol
interchangeably.}

\begin{eqnarray}
\delta \rho = - (\rho + p) \left[e^{-\Lambda} \frac{W^\prime}{r^2} + \frac{l(l+1)}{r^2} V \right] Y_{lm} -
\left(\frac{d\rho}{dr}+ \rho_{D}\frac{d\varphi}{dr}\right) e^{-\Lambda}\frac{W}{r^2} Y_{lm},
\end{eqnarray}
where in the last step we have taken into account that $a_{r}=\Phi^{\prime}$ and eq. (\ref{SSSE}).

In order to find the perturbation of the pressure we first use the
relation between the Eulerian and Lagrangian variations, namely

\begin{eqnarray}
\delta p = \Delta p - \xi^{r} \partial_{r}p
\end{eqnarray}
with $\Delta p$ being the Lagrangian variation. From the equation of
state we have

\begin{eqnarray}
\Delta p= \frac{dp}{d\rho}\Delta \rho= \frac{dp}{d\rho} \left(\delta
\rho + \xi^{r}\partial_{r}\rho\right).
\end{eqnarray}
In this way we obtain the following formula for the perturbation of
pressure

\begin{eqnarray}
\delta p=\frac{dp}{d\rho} \left\{(\rho + p)\left[e^{-\Lambda}\frac{W^\prime}{r^2} + \frac{l(l+1)}{r^2}V\right] + \rho_D
\frac{d\varphi}{dr}e^{-\Lambda}\frac{W}{r^2}\right\}Y_{lm}  - \frac{dp}{dr}e^{-\Lambda}\frac{W}{r^2}Y_{lm}.
\end{eqnarray}

Having the explicit expressions for $\delta \rho$ and $\delta p$ we
can put them into eq.(\ref{POE}) and after lengthy calculations we
obtain the dynamical equations for $W$ and $V$

\begin{eqnarray}
(\rho + p) \frac{e^{\Lambda -2\Phi}}{r^2} \partial^2_{t}W &=& \frac{d}{dr}\left\{(\rho +
p)\frac{dp}{d\rho}\left[e^{-\Lambda}\frac{W^\prime}{r^2} + \frac{l(l+1)}{r^2}V \right] +
\frac{dp}{dr}e^{-\Lambda}\frac{W}{r^2} + \frac{dp}{d\rho}\rho_{D}\frac{d\varphi}{dr} e^{-\Lambda}\frac{W}{r^2} \right\}
\nonumber \\ \nonumber \\
& - & \Phi^\prime \Big\{ (\rho + p)\left(1 + \frac{dp}{d\rho}\right)\left[e^{-\Lambda}\frac{W^\prime}{r^2} +
\frac{l(l+1)}{r^2}V \right]  \nonumber \\ \nonumber \\
&+&  \left(\frac{d\rho}{dr} + \frac{dp}{dr}\right)e^{-\Lambda}\frac{W}{r^2} + \left(1 +
\frac{dp}{d\rho}\right)\rho_{D}\frac{d\varphi}{dr}e^{-\Lambda}\frac{W}{r^2} \Big\}, \label{eq:PertW}
\\ \nonumber \\ \nonumber \\
(\rho + p)e^{-2\Phi} \partial^2_{t}V&=& (\rho + p)\frac{dp}{d\rho}\left[e^{-\Lambda}\frac{W^\prime}{r^2} +
\frac{l(l+1)}{r^2}V \right] + \frac{dp}{dr}e^{-\Lambda}\frac{W}{r^2} + \frac{dp}{d\rho}\rho_{D}\frac{d\varphi}{dr}
e^{-\Lambda}\frac{W}{r^2}. \label{eq:PertV}
\end{eqnarray}

From now on we will assume for the perturbation functions a harmonic
dependence on time, i.e. $W(r,t)=W(r)e^{i\omega t}$ and
$V(r,t)=V(r)e^{i\omega t}$. The system
(\ref{eq:PertW})--(\ref{eq:PertV}) can be considerably simplified by
combining the equations in an appropriate manner. When
differentiating equation (\ref{eq:PertV}) and adding it to equation
(\ref{eq:PertW}), and also using eq.(\ref{SSSE}), we find

\begin{eqnarray}
&&V^\prime = 2\Phi^{\prime} V - e^{\Lambda}\frac{W}{r^2} - \frac{1 +
\frac{d\rho}{dp} }{\rho + p} \rho_{D}\frac{d\varphi}{dr} V.
\end{eqnarray}
This equation together with the equation (\ref{eq:PertV}) solved for
$W^\prime$, form a system which is equivalent to
(\ref{eq:PertW})--(\ref{eq:PertV}) but much more tractable:

\begin{eqnarray} \label{FSYS}
&&W^{\prime}= \frac{d\rho}{dp}\left[\omega^2 r^2 e^{\Lambda - 2\Phi}
V + \Phi^\prime W\right] - l(l+1)e^{\Lambda} V - \frac{1 +
\frac{d\rho}{dp} }{\rho + p} \rho_{D}\frac{d\varphi}{dr} W, \nonumber\\  \\
&&V^\prime = 2\Phi^{\prime} V - e^{\Lambda}\frac{W}{r^2} - \frac{1 +
\frac{d\rho}{dp} }{\rho + p} \rho_{D}\frac{d\varphi}{dr} V.
\nonumber
\end{eqnarray}

The boundary condition at  the star surface is that the Lagrangian
perturbation of the pressure vanishes

\begin{eqnarray}\label{FBC}
(\rho + p)\left(\omega^2 e^{-2\Phi}V + \Phi^\prime
e^{-\Lambda}\frac{W}{r^2} - \frac{\rho_D}{\rho
+p}\frac{d\varphi}{dr}e^{-\Lambda}\frac{W}{r^2}\right)=0.
\end{eqnarray}

The boundary conditions at the star center can be obtained by
examining the behaviour in the vicinity of $r=0$. For this purpose
it is convenient to introduce the new functions $\tilde W$ and
$\tilde V$ defined by

\begin{eqnarray}
W= \tilde W r^{l+1}, \;\;\; V=\tilde V r^l .
\end{eqnarray}

Then one can show that at $r=0$ the following boundary condition is
satisfied

\begin{eqnarray}\label{FCAC}
\tilde W= -l \tilde V.
\end{eqnarray}

\section{Numerical results for the oscillation spectrum }
As we have shown in the previous section the properties of the MNDES
depends strongly on the dark energy content of the star and the
masses and the radii can differ significantly. It is natural to
expect that this will influence the oscillations of the star. In
order to find the oscillation spectrum of MNDES we have numerically
solved the spectral problem given by the differential equations
(\ref{FSYS}) together with the boundary conditions (\ref{FBC}) and
(\ref{FCAC}). In this section we will present the results for the
$f$-modes of the MNDES and for the higher fluid modes -- $p_1$ and
$p_2$. All the dependencies are shown up to the maximum mass for the
corresponding values of the parameters, i.e. where a change of
stability is observed.

\begin{figure}
\begin{center}
\includegraphics[width=0.6\textwidth]{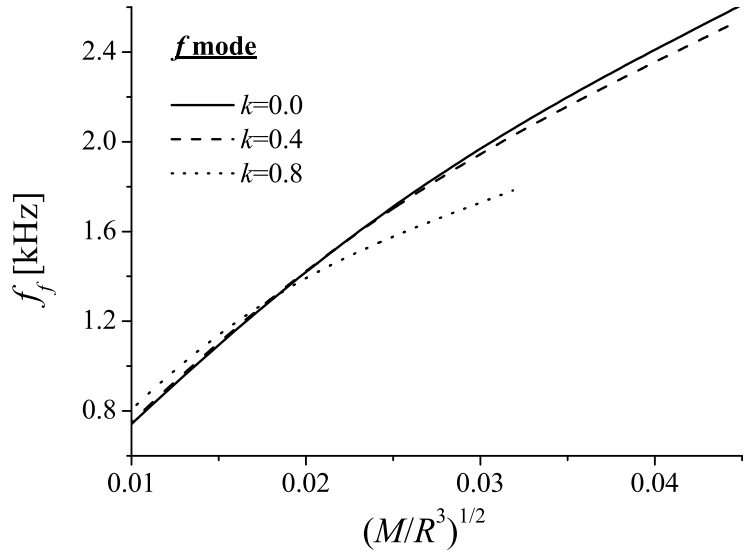}
\end{center}
\caption{The $f$-mode frequency as a function of the average density of the ordinary matter component $\sqrt{M/R^3}$.
The results
for several values of the dark energy parameter $k$ are shown.} \label{fig:f(sqrtMR3)_fmode}%
\end{figure}

For pure neutron star an empirical relation between the average
density and the $f$-mode frequency exists\footnote{The $f$-mode
frequency is proportional to the average density.} which does not
depend much on the EOS \cite{Andersson98}. As we can see from Fig.
\ref{fig:f(sqrtMR3)_fmode} this relation differ for MNDES with
different dark energy content (i.e. for different values of $k$),
but it remains relatively close to the pure neutron star (the case
with $k=0$). Only for large values of $k$ the relation changes more
significantly -- the frequencies grow slower as a function of the
average density.

\begin{figure}
\includegraphics[width=0.48\textwidth]{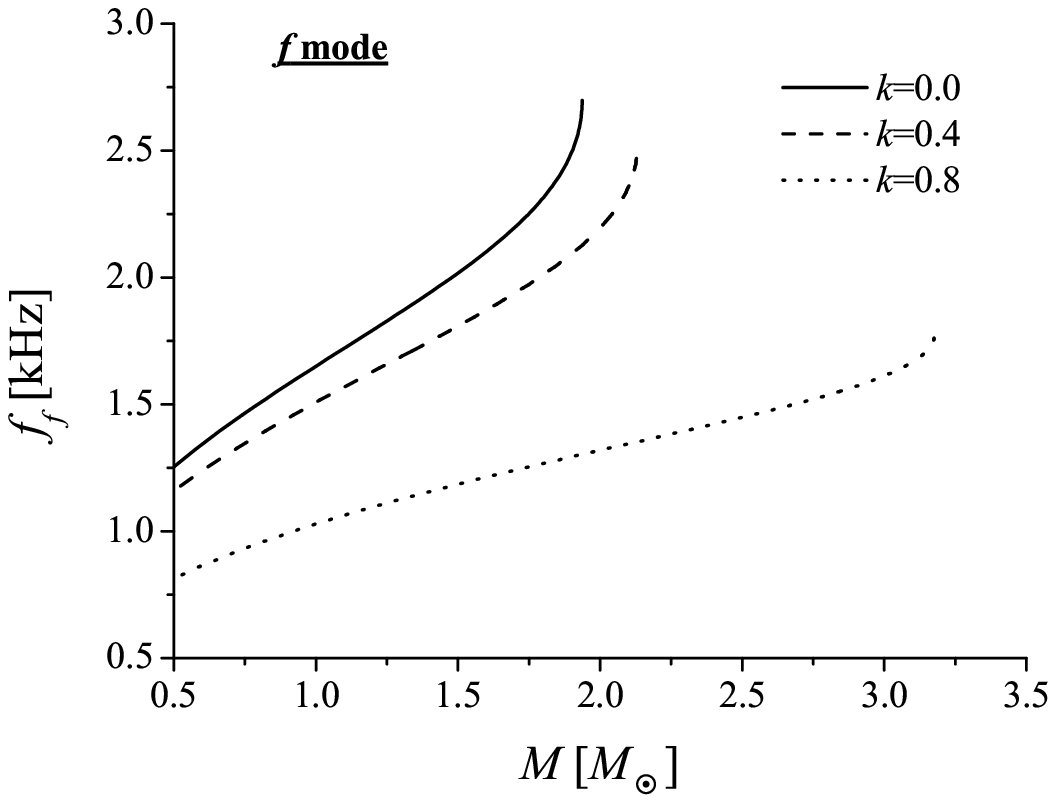}
\includegraphics[width=0.48\textwidth]{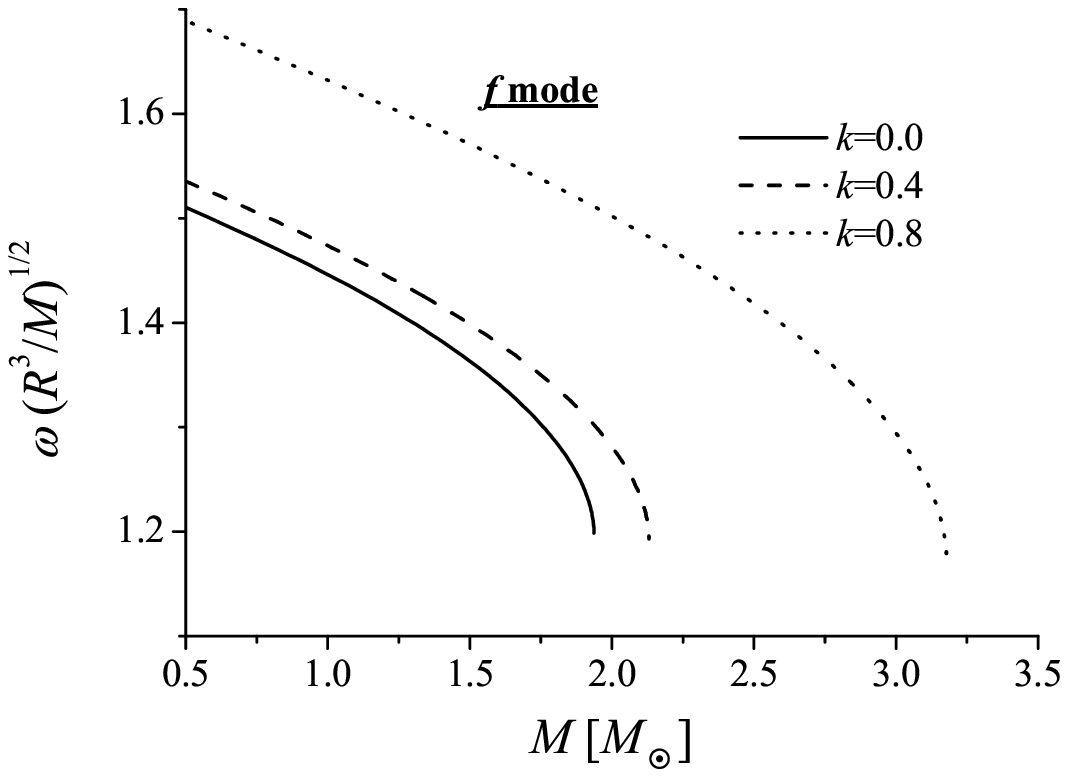}
\caption{The frequency $f$ as a function of the mass $M$ (left panel) and the normalized frequency $\omega$ as a
function of $M$ (right panel) for the $f$ mode. The results
for several values of the dark energy parameter $k$ are shown.} \label{fig:Omega(M)_fmode}%
\end{figure}

\begin{figure}
\includegraphics[width=0.48\textwidth]{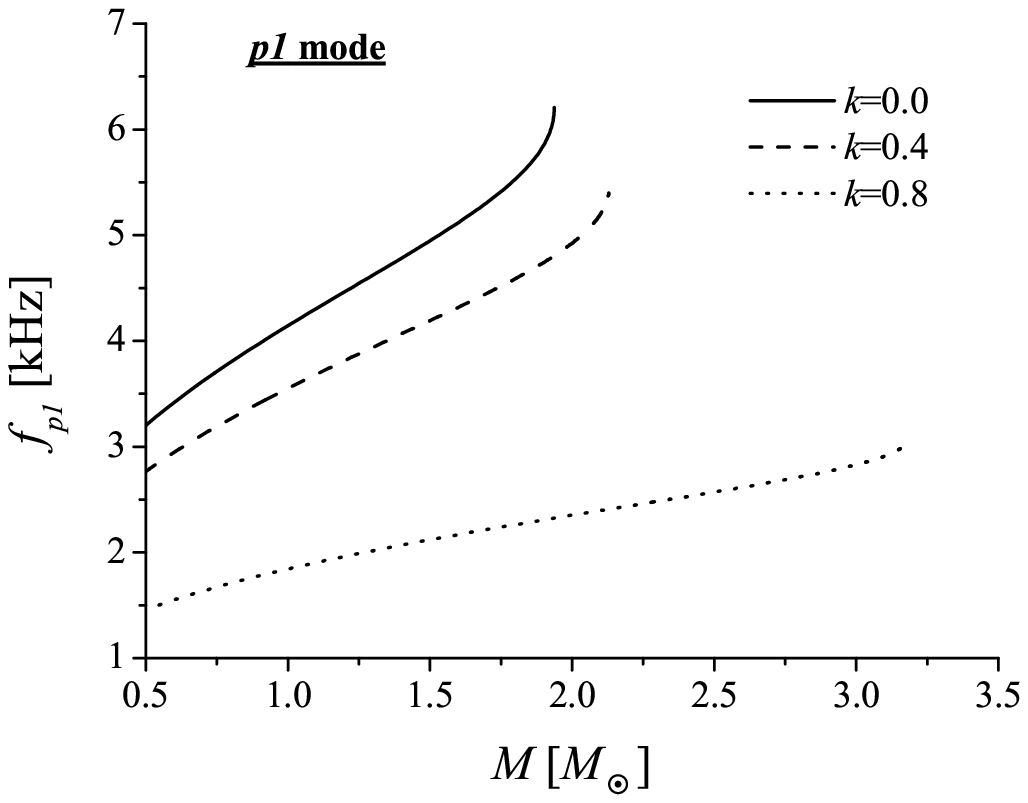}
\includegraphics[width=0.48\textwidth]{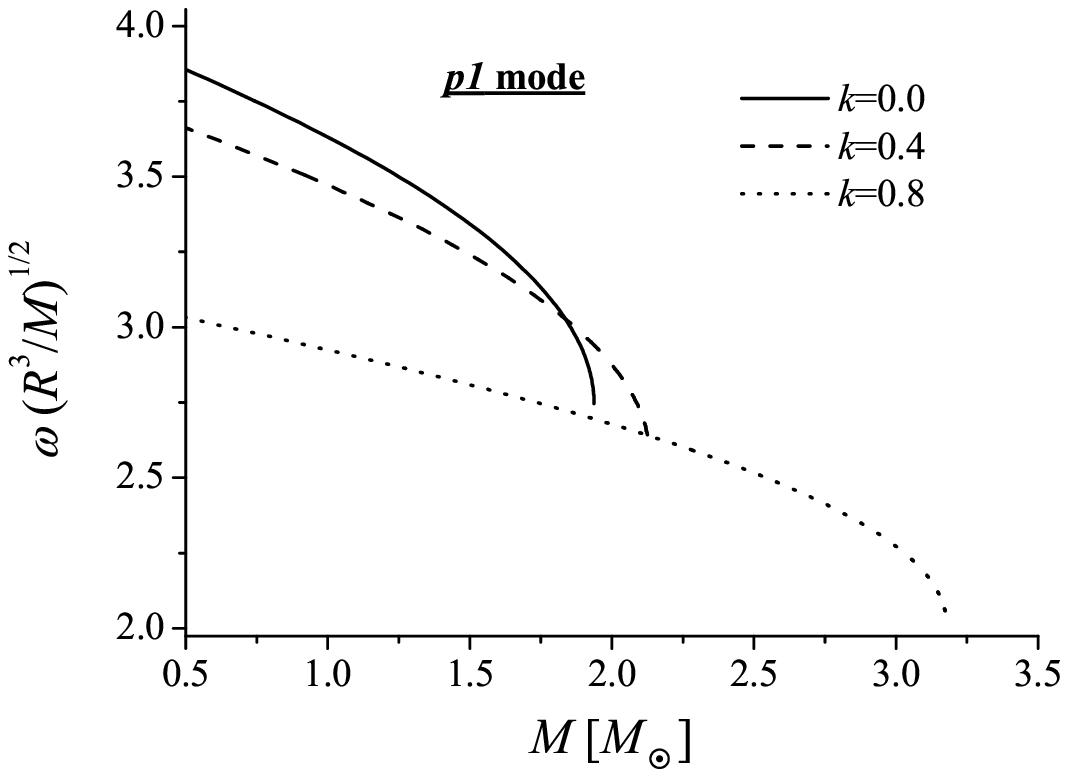}
\caption{The results for the $p_1$-mode of the same solutions as shown on Fig. \ref{fig:Omega(M)_fmode}.} \label{fig:Omega(M)_p1mode}%
\end{figure}

\begin{figure}
\includegraphics[width=0.48\textwidth]{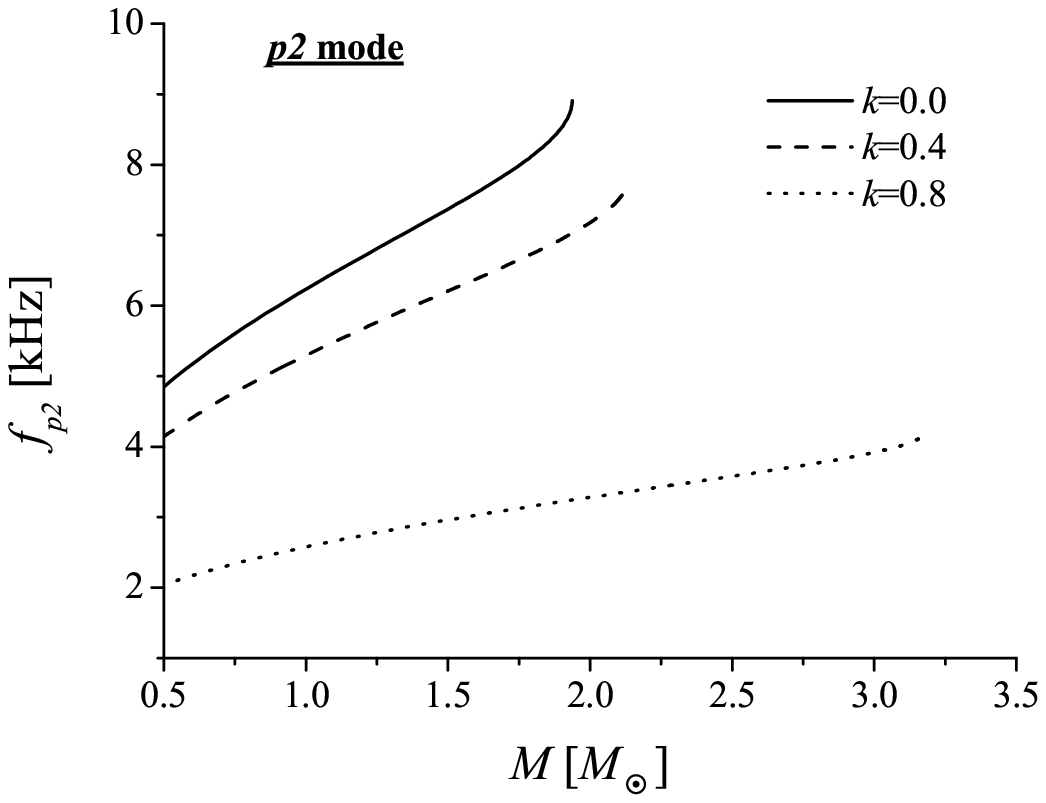}
\includegraphics[width=0.48\textwidth]{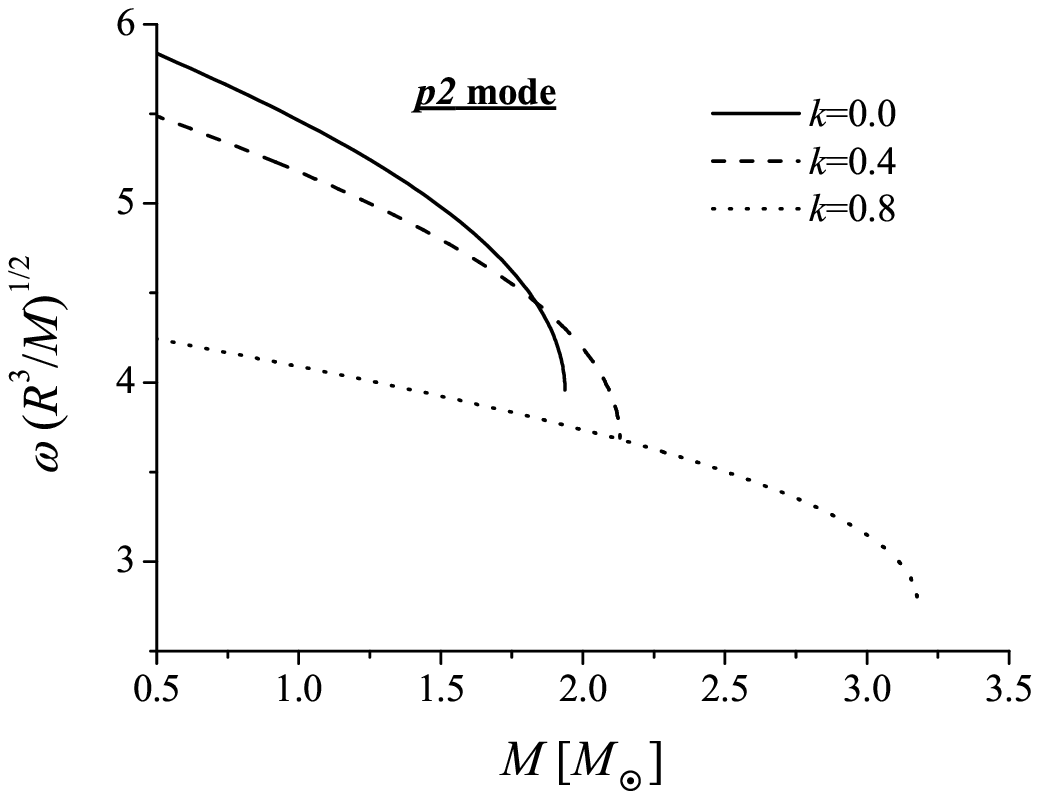}
\caption{The results for the $p_2$-mode of the same solutions as shown on Fig. \ref{fig:Omega(M)_fmode}.} \label{fig:Omega(M)_p2mode}%
\end{figure}

The influence of the dark energy on the oscillation spectrum is more
pronounced on the Figs. \ref{fig:Omega(M)_fmode},
\ref{fig:Omega(M)_p1mode} and \ref{fig:Omega(M)_p2mode} where it is
shown the oscillation frequencies $f$ and the normalized frequency
$\omega$ as a function of the total mass $M$ of the MNDES  for the
$f$, $p_1$ and $p_2$ modes. The values of the frequencies $f$
decrease when we increase the dark energy content for all of the
fluid modes and the differences with the pure neutron star case
($k=0$) can be large. This effect is more pronounced for the higher
fluid modes and for larger values of $k$ the frequencies of the
$p_2$-mode can reach two times the frequencies of the pure neutron
star case as we can see on Fig. \ref{fig:Omega(M)_p2mode}. This is a
very strong effect and by observing more than one fluid mode of the
same neutron star we can put serious constraints on the dark energy
content in the MNDES and consequently on the dark energy model
itself.

An interesting observation is that the normalized frequency $\omega$
behaves differently  as a function of the total mass $M$ for the
different fluid modes as we can see from the right panel on Figs.
\ref{fig:Omega(M)_fmode}, \ref{fig:Omega(M)_p1mode} and
\ref{fig:Omega(M)_p2mode}. The results show that the normalized
$\omega$ increases when we increase the parameter $k$ for the
$f$-mode whereas for the higher fluid modes $\omega$ decreases with
the increase of the dark energy content. This effect is
qualitatively different from most of the other alternative models of
neutron stars \footnote{Such alternative models are for example the
neutron stars in the scalar-tensor theories \cite{Sotani04} or in
the tensor-vector-scalar theory of gravity \cite{Sotani09}.} where
the qualitative behaviour of the normalized frequencies $\omega$ is
the same for all of the modes when we vary the parameters of the
corresponding model. This property can be potentially used to
distinguish between MNDES and other alternative neutron star model.
Again the differences between oscillation frequencies of the pure
neutron star and of the MNDES are larger for the higher order modes.

\begin{figure}
\begin{center}
\includegraphics[width=0.6\textwidth]{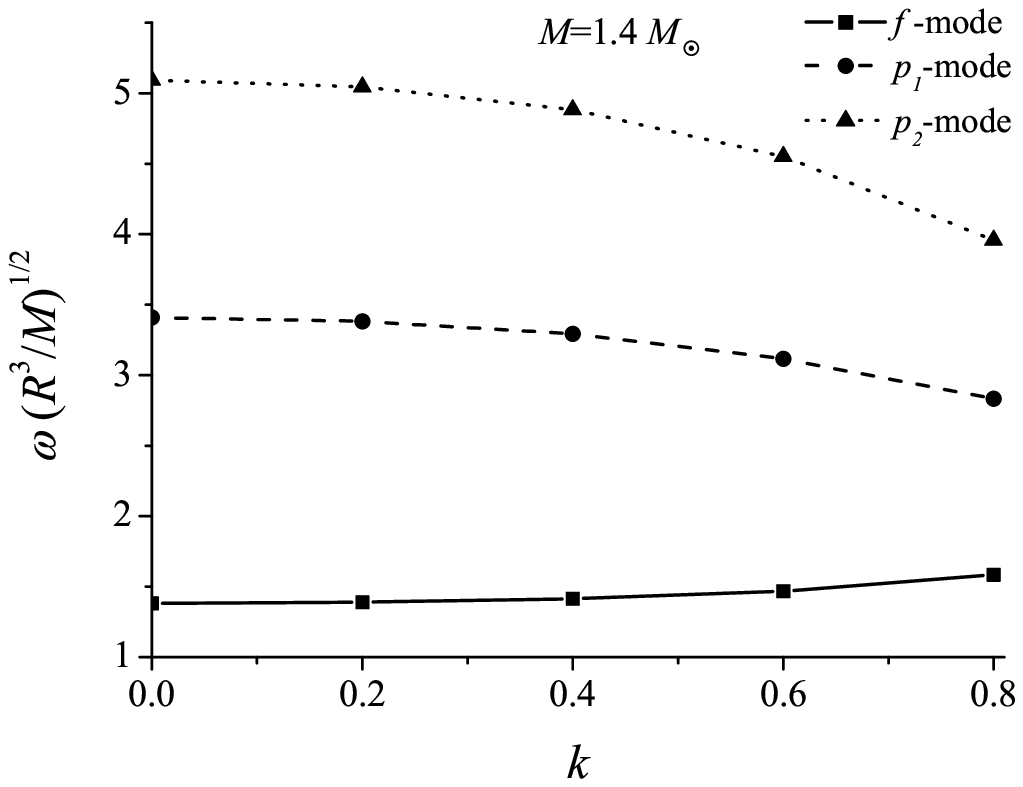}
\end{center}
\caption{The normalized frequency $\omega$ as a function of $k$ for
fixed value of the mass $M=1.4 M_{\bigodot}$. The results for the
$f$, $p_1$ and $p_2$ modes are shown.} \label{fig:Omega(kDE)_M1.4}%
\end{figure}

These properties of the oscillation spectrum can be easily seen also on Fig. \ref{fig:Omega(kDE)_M1.4} where the
normalized  frequencies of the $f$, $p_1$ and $p_2$-modes are shown as a function of $k$ for fixed value of the MNDES
mass $M=1.4 M_{\bigodot}$. The results show that when we increase the value of $k$ not only the absolute values of the
frequencies change in a bigger range for the higher order fluid modes, but also the relative change in the frequencies
is larger. On Fig. \ref{fig:Omega(kDE)_M1.4} we can also observe one of the properties of the oscillation spectrum that
we discussed above -- the normalized frequency increases when we increase the parameter $k$ for the $f$-mode whereas
for the higher fluid modes $\omega$ decreases with the increase of $k$.

\section{Conclusions}
In the present paper a model of neutron star containing dark energy
and its oscillation spectrum are studied. We model the dark energy
part of the star by using a phantom scalar field
\cite{Caldwell},\cite{CSSX},\cite{Y}. The neutron-dark-energy star
solutions are obtained by solving numerically the reduced system of
field equations. The results show that the properties of the MNDES
can be significantly different from the pure neutron star case. One
of the main difference is that the masses and the radii can be
significantly larger in the presence of dark energy. The stability
of the MNDES is also studied and it turns out that similar to the
pure neutron star case a change of stability is observed at the
critical energy density $\rho^c_0$ for which $\frac{\partial
M}{\partial \rho_{0}}=0$.

The second part of the paper is concentrated on examining the
possible dark energy imprints in gravitational wave spectrum of the
obtained MNDES. In order to do that we study the oscillation
spectrum of the obtained solutions in the Cowling approximation
where the non-fluid degrees of freedom are kept fixed. The values of
the oscillation frequencies of the MNDES can differ significantly
from the pure neutron star case and the differences are bigger for
the higher fluid modes. Thus the observation of gravitational waves
emitted from neutron stars could be used to detect the presence of
dark energy in the neutron stars  and to put constraints on the dark
energy properties in general.

We will finish by mentioning a possible further extension of the
present work. Since the scalar phantom field violates the weak/null
energy condition it can maintain configurations with nontrivial
topology -- the so-called wormholes. Such a solution describing a
star harbouring a wormhole at its core was recently found in
\cite{DFKK}. It would be very interesting to study the influence of
the possible nontrivial topological structures in neutron stars,
like wormholes, on the gravitational wave spectrum of the stars.

\vspace{1.5ex}
\begin{flushleft}
\large\bf Acknowledgments
\end{flushleft}

The authors would like to thank K. Kokkotas for reading the
manuscript and for the valuable suggestions.  S.Y. would like to
thank the Alexander von Humboldt Foundation for the support, and the
Institut f\"ur Theoretische Astrophysik T\"ubingen for its kind
hospitality. D.D. would like to thank the DAAD for the support and
the Institut f\"ur Theoretische Astrophysik T\"ubingen for its kind
hospitality. D.D. is also supported by the Transregio~7
``Gravitational Wave Astronomy'' financed by the Deutsche
Forschungsgemeinschaft DFG (German Research Foundation). This work
was also supported in part by  the Bulgarian National Science Fund
under Grants DO 02-257 and DMU-03/6.


\begin{thebibliography}{99}

\bibitem{R1} A. G. Riess et al. [Supernova Search Team Collaboration], Astron. J. {\bf 116}, 1009 (1998);
 S. Perlmutter et al. [Supernova Cosmology Project Collaboration], Astrophys. J. {\bf 517}, 565 (1999);
 P. Astier et al. [The SNLS Collaboration], Astron. Astrophys. {\bf 447}, 31 (2006);
 A. G. Riess et al. [Supernova Search Team Collaboration], Astrophys. J. {\bf 607}, 665 (2004);
 A. G. Riess et al., Astrophys. J. {\bf 659}, 98 (2007);   N. Spergel et al. [WMAP Collaboration], Astrophys. J. Suppl. 170, 377 (2007);
 M. Kowalski et al. [Supernova Cosmology Project Collaboration], Astrophys. J. 686, 749 (2008);
 E. Komatsu et al. [WMAP Collaboration], Astrophys. J. Suppl. 180, 330 (2009).

\bibitem{ASSS} U. Alam, V. Sahni, T.D. Saini and A.A. Starobinsky, Mon. Not. R. Astron. Soc. {\bf 354}, 275 (2004).


\bibitem{MM} P. Mazur, E. Mottola, [arXiv:gr-qc/0405111]

\bibitem{Lobo} F. Lobo, Class. Quant. Grav. {\bf 23}, 1525 (2006).


\bibitem{BF} K. Bronnikov, J.C. Fabris, Phys. Rev. Lett. {\bf 96},  251101 (2006).

\bibitem{DFMS} V. Dzhunushaliev, V. Folomeev, R. Myrzakulov, D. Singleton, JHEP {\bf 0807}, 094 (2008).


\bibitem{CdSvdR} R. Chan, M.F.A. da Silva, J.F. Villas da Rocha,  Gen. Rel. Grav. {\bf 41}, 1835 (2009).

\bibitem{Ghezzi} C. Ghezzi, [arXiv:0908.0779[gr-qc]]


\bibitem{RYMS} F. Rahaman, A. Yadav, S. Ray, R.Maulick, R. Sharma, [arXiv:1102.1382[gr-qc]]


\bibitem{DFKK} V. Dzhunushaliev, V. Folomeev, B. Kleihaus, J. Kunz, JCAP {\bf 1104} 031
(2011).

\bibitem{Y} S. Yazadjiev, Phys. Rev. {\bf D83}, 127501 (2011),



\bibitem{Andersson98} N. Andersson and K. D. Kokkotas, Mon. Not. R. Astron. Soc. \textbf{299}, 1059 (1998).

\bibitem{Kokkotas01} K. D. Kokkotas, T. A. Apostolatos, and N. Andersson, Mon.
Not. R. Astron. Soc. \textbf{320}, 307 (2001).


\bibitem{Caldwell} R. R. Caldwell,  Phys. Lett. {\bf B 545}, 23 (2002).

\bibitem{CSSX} Yi-Fu Cai, E. Saridakis, M. Setare, J.-Q. Xia, Phys. Rept. {\bf 493}, 1 (2010)




\bibitem{PT} F. Piazza, S. Tsujikawa, JCPA {\bf 0407}, 004 (2004).

\bibitem{NO} S. Nojiri, S. Odintsov, Phys. Lett. {\bf B562}, 147 (2003).

\bibitem{CHT} S. Carroll, M. Hoffman, M. Troden, Phys. Rev. {\bf D 68}, 023509 (2003).

\bibitem{BCCF} K.A. Bronnikov, G. Clement, C.P. Constantinidis, J.C. Fabris,  Phys. Lett.  {\bf A243}, 121 (1998).

\bibitem{APicon} C. Armend´ariz-Picon, Phys. Rev. {\bf D 65}, 104010 (2002).

\bibitem{Sen1} A. Sen, JHEP {\bf 0204}, 048 (2002);   JHEP {\bf 0207}, 065 (2002)


\bibitem{GPV} M. Gasperini, F. Piaza, G. Veniziano, Phys. Rev. {\bf D 65}, 023508 (2001).

\bibitem{KKLP} N. Khviengia, Z. Khviengia, H. L\"u, C. Pope, Class. Quant. Grav. {\bf 15}, 759 (1998).

\bibitem{Nilles} H. Nilles, Phys. Rep. {\bf 110}, 1 (1984).

\bibitem{Percival} Percival, W.J., et al., MNRAS {\bf 401}, 2148 (2010)
\bibitem{Komatsu11} E. Komatsu et al., Astrophys. J. Suppl. \textbf{192}, 18 (2011)
\bibitem{Blacke} Blake, C., et al., MNRAS {\bf 415}, 2892 (2011)
\bibitem{Beuter} Beutler, F., et al., 2011, arXiv:1106.3366,
\bibitem{Escamilla} Escamilla-Rivera, C.; et al., JCAP {\bf 09}, 003
(2011)
\bibitem{Li} Li, X.-D., et al., JCAP {\bf 07},  011 (2011).


\bibitem{Herrera} L. Herrera and N. O. Santos, Phys. Rep. {\bf 286}, 53 (1997).


\bibitem{Alonso85} J. Diaz-Alonso and J.M. Iba\~nez-Cabanell, Astrophys. J. \textbf{291}, 308 (1985).






\bibitem{McDermott} McDermott P.N., Van Horn H.M., Scholl J.F., 1983, ApJ, 268, 837

\bibitem{Lindblom} Lindblom L., Splinter R.J., 1990, ApJ, 348, 198

\bibitem{Yoshida} S. Yoshida, Y. Kojima, Mon. Not. R. Astron. Soc. \textbf{289}, 117 (1997)







\bibitem{Sotani04} H. Sotani and K. D. Kokkotas, Phys. Rev. \textbf{D70}, 084026 (2004)

\bibitem{Sotani09} H. Sotani, Phys. Rev. \textbf{D79}, 064033 (2009)



\end{thebibliography}
\end{document}